\documentclass{article}
\usepackage{geometry} 
\usepackage{amsmath}
\usepackage{array}
\usepackage{graphicx}
\usepackage[pdftex]{hyperref}
\usepackage[usenames]{color}
\usepackage{fancyhdr}
\usepackage{verbatim}
\usepackage{setspace}
\usepackage{graphicx}
\usepackage{tikz}
\usepackage{pgfmath}
\usepackage{amsfonts}

\newcommand{\xb}{\mathbf{x}}
\newcommand{\yb}{\mathbf{y}}
\newcommand{\ub}{\mathbf{u}}
\newcommand{\Hb}{\mathbf{H}}
\newcommand{\myparsum}[1]{}
\definecolor{dgreen}{rgb}{0,0.5,0.0}
\usetikzlibrary{arrows,decorations.pathmorphing,backgrounds,positioning,fit} 

\title{\textbf{A framework for simulating and estimating the state and functional topology of complex dynamic geometric networks}}
\author{Marius Buibas\\ \textit{Department of Bioengineering}\\ \textit{University of California, San Diego}\and \\Gabriel A. Silva\\ \textit{Departments of Bioengineering and Ophthalmology}\\ \textit{Neurosciences Program}\\ \textit{University of California, San Diego}}

\date{}

\begin{document}
\maketitle{}
{\noindent \textbf{Abstract-}
We present a framework for simulating signal propagation in geometric networks (i.e. networks that can be mapped to geometric graphs in some space) and for developing algorithms that estimate (i.e. map) the state and functional topology of complex dynamic geometric networks.  Within the framework we define the key features typically present in such networks and of particular relevance to biological cellular neural networks: Dynamics, signaling, observation, and control.  The framework is particularly well-suited for estimating functional connectivity in cellular neural networks from experimentally observable data, and has been implemented using graphics processing unit (GPU) high performance computing. Computationally, the framework can simulate cellular network signaling close to or faster than real time. We further propose a standard test set of networks to measure performance and compare different mapping algorithms.}\\\

\noindent \rule{6.25in}{0.5pt}
Address correspondance to Dr. Gabriel A. Silva, UC San Diego Jacobs Retina Center, 9415 Campus Point Drive, La Jolla, CA 92037-0946 USA. Electronic mail: gsilva@ucsd.edu

\newpage
\section{Introduction}

Complex dynamic networks permeate many real world engineering and biological applications. 
The development of mathematical and computational tools for understanding and predicting network dynamics will be key to manipulating and interacting with such real world networks. 
Network theory is a subset of graph theory where the connections between vertices have a number value describing some attribute of that connection, such as for example bandwidth, flow rates, or a cost function.  Complex networks are defined to have a non-standard topology, i.e. the functional links between nodes in the network, implying some structure in the connectivity pattern of the network beyond a simple lattice or complete random connectivity.  Biological cellular neural networks are both complex and dynamic, meaning that the connection attribute between any two vertices may change with respect to time and, more importantly, individual vertices exhibit their own nonlinear signaling dynamics. Complex functional interactions of networks made up of large numbers of neurons and glia produce emergent systems-level phenomena such as consciousness and self-awareness, and are responsible for how neural information is represented and processed.  
Changes in the structure of such networks presumably underlie the development of multidimensional central nervous system disorders. 
For example, hypersynchronous neuronal and glial activity in networks of neurons are associated with the paroxysmal depolarization shifts that underlie epilepsy \cite{Tian:2005p1755,Sudbury:2007p2527,Wetherington:2008p2523,Feldt:2007p2540}. Ultimately, the physiologic behavior of a neural cell network is dependent on both its functional topology and the dynamics of individual cells. 


Within a complex dynamic network there are two topologies.  A static, structural topology that describes all the possible connections within the network, and a dynamic, functional topology that establishes how a signal propagates through the static topology.  Functional topologies are subsets of the structural topology and vary depending on the functional connectivity, internal dynamics of individual vertices, and the specific stimulus to the network.  While this is the case for biological neural networks, where cells that are physically connected need not necessarily signal each other, in cellular neural circuits and networks structure and function influence each other and the states of cells and the connections between them may change with time as a function of plasticity mechanisms. However, structural changes in the physical connectivity of a cellular neural network leading to changes in the connectivity topology occur on a very different time scale then functional changes that can be influenced relatively quickly by plasticity mechanisms that produce changes in signaling efficacy between cells (i.e. changes in connectivity weights). While the observation of the structural network topology of cellular neural networks may be experimentally very challenging (and indeed is the focus of much intense research), it is a relatively straightforward task. The observation of functional topologies in biological neural networks however poses additional experimental and theoretical challenges that need to be considered.  Signaling events and resultant networks may be unique and be observable only once as a signal propagates through a network.  The functional topology is dynamic and may change during observation.  Noise and unknown external factors limit observability and reduce repeatability. These factors make the estimation of functional connectivity from observed activity a difficult task, though a critical one for systems neuroscience if we are to understand how dynamic functional signaling in the brain at the level of networks and circuits produces responses and behaviors in the organism. 

Current approaches for studying cellular neural networks can be roughly classified into three categories.  The first and most popular amongst experimentalists are statistical methods that correlate the activities of two or more neurons in a network. This provides purely descriptive statistics about the behavior of cells. For the most part, statistical approaches make no underlying assumptions about the cellular and systems dynamics that give rise to observed signals in a network of cells.  Another way to study networks is through simulation of networks with known connectivities and dynamic parameters in order to simulate real-world observed system level phenomena such as vision and audition. Using well established environments like NEURON or Genesis,  many real-world phenomena have been described through simulation. However, dynamic parameters and functional connections are manually specified in simulation environments such as these in order to achieve results that mimic biological function, requiring the estimation of experimentally unobservable variables.  The third category is in some ways the reverse process to simulation, where temporal data is used with appropriate models in order to estimate parameters. Within this third category, we introduce a modeling framework for using real-world data to map the functional topology of complex dynamic networks. While not a mapping algorithm or simulation environment, the framework formally defines key features of cellular neural network signaling and experimental constraints associated with observation and stimulus control, and can accommodate any appropriate model of intracellular dynamics.  Alongside the definition of the framework, a test set of synthetic networks with known connectivities is provided to help the development of mapping algorithms by providing a common benchmark any such algorithm should be able to map. In 
a subsequent paper to this one we will introduce an approach that will estimate and map the functional 
topology of complex networks with unknown connectivities given limited 
and often noisy observations that takes advantage of the results introduced here.
 
The proposed framework has a number of unique properties that makes it particularly applicable to the constraints and experimental limitations imposed by real biological cellular neural networks.  First, dynamic activity and signaling is modeled at the individual node (i.e. cell) scale. The dynamics of individual cells are modeled as state sets, with transition functions describing their evolution across discrete time steps. Cellular resolution was chosen because it represents the best compromise between observability, dynamics, and complexity.  Large numbers of individual cells can now be observed in parallel in functional neural networks using optical microscopy \cite{Homma:2009p1581, Benninger:2008p1618, Garaschuk:2006p1712, Nikolenko:2007p572}.  Single cell neuronal dynamics are well understood and many models exist (see for example \cite{Dayan:2001p3694, Trappenberg:2009p3744}), while similar models of single cell astrocyte dynamics are beginning to emerge \cite{Stamatakis:2006p1199, Nadkarni:2008p3684, DePitta:2009p1741, Bennett:2005p990, Lavrentovich:2008p1531}.  Attempting to go to a finer, sub-cellular compartmental resolution dramatically increases the complexity of the model, computational demand, and is generally not experimentally observable at a network level. 
Secondly, cells are located in physical space and their positions are easily determinable during experimental observation. When connected cellular networks form geometric networks.
Thirdly, the effect of a signal on a target cell is defined as a state change in the target cell in response to the influence of a source cell that connects to it. That influence is not instantaneous, and is delayed by the physical distance between cells and the speed of transmission.  Signals are modulated in strength by functional weights, which establish the magnitude of the influence.
Fourthly, to more realistically simulate experimental conditions and measurements, noise can be added to multiple levels within the framework, from parameters to state and observation variables.
Finally, experimental user-defined controls at the individual cell level are defined within the framework.  Controls should be designed to make observations more informative of the network dynamics, but should not change the underlying parameters and connectivities.  The framework is described in detail in section \ref{sec:framework}. The results section (section \ref{sec:results}) shows how single cell dynamic models are integrated within the framework (\ref{sec:indyn}), and how network connectivity is established from individual cells (\ref{sec:netsig}). We also describe how the framework accommodates plasticity mechanisms (\ref{sec:plasticity}) and experimental observability associated with optical calcium imaging (\ref{sec:observ}). Section \ref{sec:impl} discusses the practical implementation of the framework using high performance graphical processing unit (GPU) computing. 

In section \ref{sec:tests} we use the framework to propose a standard set of benchmark test networks of varying sizes and topologies to evaluate and compare different network mapping algorithms.  Mapping algorithms would have access to simulated observable data (i.e. simulated experimental data) generated by the framework as a function of chosen test networks and be required to derive the unobservable parameters and functional network connectivity.  The concept of a standardized test to gauge the effectiveness of an algorithm is not new, especially for optimization algorithms.  For example, in the field of nonlinear programming and optimization a standard benchmark set was established in a landmark collection of test problems \cite{Hock:1980p4454} that are used for testing any nonlinear optimization algorithm.  Test collections have grown and developed into problem environments, providing the underlying problem code to be used directly by the optimizers \cite{Bondarenko:1999p4732,GOULD:1995p4733}). By providing a set of problems with known solutions, algorithm developers have a standard by which to measure solution accuracies, convergence rates, computation times, and suitability to different problem types. We propose that a similar test set for algorithms designed to identify and map functional cellular neural networks and circuits will be just as useful.  To address this, we have developed computer code that generates observable data from a known network and connectivity.  The code encompasses all the elements of the framework, runs in real time for all the test networks, and is designed for parallel computation, and can therefore be used as a starting point for mapping algorithms. 

\section{A Framework for Dynamics, Signaling, Control and Observation in Geometric Networks} \label{sec:framework}
We develop the proposed framework using standard graph theoretic and set theoretic concepts and terminology. In the most general sense, a network is a type of graph.  A graph is defined as an ordered pair of finite disjoint sets $(v,E)$ such that  $\mathbf{v}$ is the set of $J$ vertices of $G$ and $\mathbf{E}$ is the set of edges of $G$, i.e. $\mathbf{v}=v(G)$ is the vertex set of $G$ while $\mathbf{E}=E(G)$ is the edge set of $G$. An edge $e_{ij}$ is defined if there is a directed connection from vertex $i$ to vertex $j$. Geometric graphs are graphs where the relative positions of vertices are assigned coordinates in some geometric space. While this is the most generic description of a graph, dynamic geometric networks as we use the term here are more specialized cases of generalized geometric graphs defined as follows. Vertices in a network have two attributes, a known and static position in physical Cartesian space denoted by $\xb_j$ for a given vertex $j$ and a time-variant state set $\yb_{j}(t)$ of $K_j$ state variables:
\begin{subequations}
\begin{equation}\label{eq:yjstate}
\yb_{j}(t)=\{ y_{1,j}(t),y_{2,j}(t),\dots,y_{K_j,j}(t)\}
\end{equation}
such that formally
\begin{equation} \label{eq:yjstateformal}
\yb_j (t)=\{y_{k,j}(t): k \in \mathbb{N}, k \le K_j\} \text{ for any given vertex $j$}
\end{equation}
\end{subequations}

Next, for all vertices $i$ other than $j$, let the set $\mathbf{Y}_{j}(t)$ be the union of all $i$, i.e. the collection of states of all vertices in the network excluding vertex $j$, weighted and delayed relative to vertex $j$, in the sense that every vertex $i$ has the potential to pass information (e.g. a signal) to vertex $j$ with varying amounts of 'influence' as determined by a collection of weights that modulate any directed edges from $i$ to $j$. Furthermore, such information will be delayed by some finite time as a function of the geometric position of vertex $i$ in the network relative to $j$ and the finite speed of information propagation. We define 
\begin{equation}\label{eq:Yjset}
\mathbf{Y}_{j}(t)=\cup_{i\in \mathbb{N};i\le J;i \ne j} \Omega_{ij}(t)\cdot \yb_{i}(t-\tau_{ij})\}
\end{equation}
where with out loss of generality we define
\begin{equation}
\Omega_{ij}(t)=[\omega_{1,ij}(t),\omega_{2,ij}(t),\dots,\omega_{K,ij}(t)]
\end{equation}
and restrict \ref{eq:yjstateformal} for vertex $i$ with temporal delays as vector sets, i.e. 
\begin{equation} \label{eq:yistateformal}
\yb_i (t)=[y_{k,i}(t): k \in \mathbb{N}, k \le K_i] \text{ for any given vertex $i \ne j$}
\end{equation}
The delays $\tau_{ij}$ are non-negative values representing the delay of information passing from $i$ to $j$. In all cases, here and below we adopt the convention that indexing subscripts given by '$ij$' enumerate the variable that uses the subscript as linking vertex pairs $i$ and $j$.

We then define a transition function $\mathbf{H}_{j}(\cdot)$ with parameter set $\Theta_{j}$ that describes the temporal progression or evolution of $\yb_{j}(t)$ in discrete time increments $\Delta t$:
\begin{equation}\label{eq:Htrans}
\yb_{j}(t+\Delta t)=\mathbf{H}_{j}\big(\yb_{j}(t),\mathbf{Y}_{j}(t),\ub_{j}(t),\Theta_{j}\big) 
\end{equation}
where $\mathbf{H}_{j}(\cdot)$ is given by 
\begin{equation}\label{eq:Hj}
\mathbf{H}_{j}=\cup_{k\in \mathbb{N};k\le K_j}H_{k,j}(\yb_{j}(t),\mathbf{Y}_{j}(t),u_{k,j}(t),\Theta_{k,j})
\end{equation}
$\ub_{j}(t)$ is a user control or experimental input.
\begin{equation}
\ub_{j}(t)=\cup_{k\in \mathbb{N};k\le K_j}u_{k,j}(t) 
\end{equation}
and $\Theta_{j}$ is parameter set 
\begin{subequations}
\begin{equation}
\Theta_{k,j}=\{\theta_{l,k}: l, k \in \mathbb{N}; l \le L_k; k \le K_j\}
\end{equation}
\begin{equation}
\Theta_j=\cup_{k \in \mathbb{N};k \le K_j} \Theta_{k,j}=\cup^{K_j}_{k=1} \Theta_{k,j} \text{ for any given vertex $j$}
\end{equation}
\begin{equation} \label{eq:allparams}
\text{and } \mathbf{\Theta}_J=\cup_{j \in \mathbb{N};j \le J} \Theta_j=\cup^J_{j=1} \Theta_j
\end{equation}
\end{subequations}
$L_k$ is the number of parameters for a given state variable, $K_j$ is the number of state variables for a given vertex $j$, and $J$ represents the size of the network (i.e. the total number of vertices).  Note that the functions comprising the set $\mathbf{H}_{j}(\cdot)$, each advance their respective variables in time:
\begin{align*}
y_{1,j}(t+\Delta t)&=H_{1,j}(y_{1,j}(t),\mathbf{Y}_{j}(t),u_{1,j},\Theta_{1,j}) \\
y_{2,j}(t+\Delta t)&=H_{2,j}(y_{2,j}(t),\mathbf{Y}_{j}(t),u_{2,j},\Theta_{2,j}) \\
\cdots \\
y_{K,j}(t+\Delta t)&=H_{K_j,j}(y_{K_j,j}(t),\mathbf{Y}_{j}(t),u_{K_j,j},\Theta_{K_j,j}) 
\end{align*}

Similarly, we define a function $\mathbf{G}_{ij}(\cdot)$ that describes the time course of the weighing sets $\Omega_{ij}(t)$ with parameter sets $\Lambda_{ij}$ as follows:
\begin{equation}\label{eq:omegatrans}
\Omega_{ij}(t+\Delta t)=\mathbf{G}_{ij}\big(\Omega_{ij}(t),\yb_{j}(t),\yb_{i}(t-\tau_{ij}),\Lambda_{ij}\big)
\end{equation}
where analogous with \ref{eq:Hj} $\mathbf{G}_{ij}(\cdot)$ is given by 
\[ \mathbf{G}_{ij}=\cup_{k\in \mathbb{N};k\le K}G_{k,j}(\Omega_{ij}(t),\yb_{j}(t),\yb_{i}(t-\tau_{ij}),\Lambda_{k,ij}\big) \]
with parameters
\begin{subequations}\label{eq:Gparams}
\begin{equation}
\Lambda_{k,ij}=\{\lambda_{l,k}: l, k \in \mathbb{N}; l \le L_k; k \le K\}
\end{equation}
\begin{equation}
\mathbf{\Lambda_{ij}}=\cup_{k \in \mathbb{N};k \le K} \Lambda_{k,ij} \text{ for any given vertex pair $ij$}
\end{equation}
\end{subequations}

The delays between vertex pairs $\tau_{ij}$ are defined as functions of the positions of the two vertices:
\begin{equation}\label{eq:gendelay}
\tau_{ij}=D\big(\xb_{i},\xb_{j},\Gamma_{ij}\big);\quad D(\cdot)\geq 0
\end{equation}
where the set $\Gamma_{ij}$ is the set of parameters of the non-negative function $D$, specific to the pair $ij$.  

Formally, the temporal evolutions of $\yb_{j}(t)$ and $\Omega_{ij}(t)$ are continuous and expressed as a discrete delay differential equations with delays $\tau_{ij}$ for all vertices connecting to vertex $j$, so that the continuous forms of equations \ref{eq:Htrans} and \ref{eq:omegatrans} are 
\begin{equation}\label{eq:ycont}
\frac{\partial \yb_{j}(t)}{\partial t}=\mathbf{H}_{j}\big(\yb_{j}(t),\mathbf{Y}_{j}(t),\ub_{j}(t),\Theta_{j}\big) 
\end{equation}
and 
\begin{equation}\label{eq:omegacont}
\frac{\partial \omega_{ij}(t)}{\partial t}=\mathbf{G}_{ij}\big(\Omega_{ij}(t),\yb_{j}(t),\yb_{i}(t-\tau_{ij}),\Lambda_{ij}\big)
\end{equation}
In the limit as $\Delta t \rightarrow 0$ \ref{eq:Htrans} and \ref{eq:omegatrans} can be written as $\yb_{j}(t+dt)$ and $\Omega_{ij}(t+dt)$ and equations \ref{eq:ycont} and\ref{eq:omegacont} apply. However, from a practical experimental perspective time measurements will always be finite and the discrete forms need be considered. As such, in this paper we do not pursue further the interesting theoretical implications of the continuous forms given by \ref{eq:ycont} and\ref{eq:omegacont}.

Finally, we define an observation set $\mathbf{z}_{j}(t)$ composed of $M$ variables, that operates directly on the state set $\yb_{j}(t)$:
\begin{equation}\label{eq:obsfun}
\mathbf{z}_{j}(t)=\mathbf{F}_{j}\big(\yb_{j}(t),\Phi_{j}\big)
\end{equation}
The observation function $\mathbf{F}_{j}$ is vector valued:
\begin{equation}
\mathbf{F}_{j}(t)=\cup_{m\in \mathbb{N};m\le M}F_{m,j} \big(\yb_{j}(t),\Phi_{m,j}\big)
\end{equation}
with parameter set $\Phi_{j}$ given by
\begin{subequations}
\begin{equation}
\Phi_{m,j}=\{\phi_{l,m}: l, l \in \mathbb{N}; l \le L_m; m \le M\}
\end{equation}
\begin{equation}
\mathbf{\Phi_j}=\cup_{m \in \mathbb{N};m \le M} \Phi_{m,j}\text{ for any given vertex $j$}
\end{equation}
\end{subequations}


The framework presented here is general, as it allows for communication between any two state variables between any two vertices.  Transition functions and their parameters are defined specific to vertex $\mathbf{H}_{j}(\cdot)$ or communication between vertex pairs $\mathbf{G}_{ij}(\cdot)$.  This produces a large set of functions and parameters, though in practice one or two different functions are applied to all cells or combinations.  The weighing set $\Omega_{ij}$ can operate on all state variables of the connecting vertex $i$ into target $j$, though usually one state of $i$ is transmitted to one state in $j$.  In the next section, we will describe how several dynamics and communication models used in cellular networks fit within this framework to reproduce observable quantities similar to experimentally measured data.  

\section{Results}\label{sec:results}
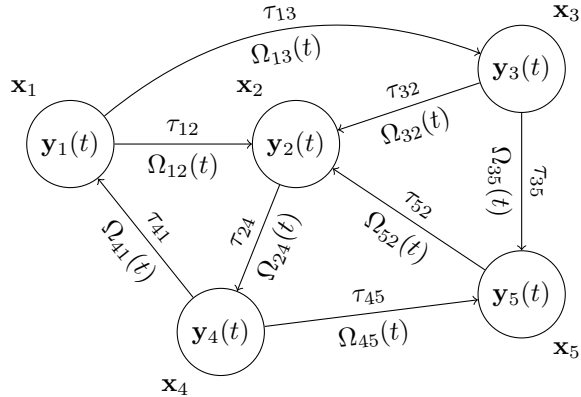
\begin{figure}[htbp]
\begin{center}
\begin{tikzpicture}
	\node [draw,circle,label=120:$\xb_{1}$] (1) at (0,3) { $\yb_{1}(t)$};
	\node [draw,circle,label=120:$\xb_{2}$] (2) at (3,3) { $\yb_{2}(t)$};
	\node [draw,circle,label=60:$\xb_{3}$] (3) at (6,4) { $\yb_{3}(t)$};
	\node [draw,circle,label=-120:$\xb_{4}$] (4) at (2,0.5) { $\yb_{4}(t)$};
	\node [draw,circle,label=-60:$\xb_{5}$] (5) at (6,1) { $\yb_{5}(t)$};
	\path[->]  
		(1) edge node[above] {$\tau_{12}$} node[below] {$\Omega_{12}(t)$} (2)
		(1) edge [bend left] node[sloped,above] {$\tau_{13}$} node[sloped,below] {$\Omega_{13}(t)$} (3)
		(3) edge node[sloped,above] {$\tau_{32}$} node[sloped,below] {$\Omega_{32}(t)$} (2)
		(2) edge node[sloped,above] {$\tau_{24}$} node[sloped,below] {$\Omega_{24}(t)$} (4)
		(4) edge node[sloped,above] {$\tau_{45}$} node[sloped,below] {$\Omega_{45}(t)$} (5)
		(4) edge node[sloped,above] {$\tau_{41}$} node[sloped,below] {$\Omega_{41}(t)$} (1)
		(3) edge node[sloped,above] {$\tau_{35}$} node[sloped,below] {$\Omega_{35}(t)$} (5)
		(5) edge node[sloped,above] {$\tau_{52}$} node[sloped,below] {$\Omega_{52}(t)$} (2);
\end{tikzpicture}
\caption{A five-vertex dynamic network.  Each vertex $j$ has a position in physical Cartesian space denoted by vector $\xb_{j}$, and a dynamic state set $\yb_{j}(t)$.  A vertex's dynamic state varies in discrete time steps, and is influenced by its own previous state and the states of other vertices connecting into it, with a delay $\tau$ and a functional connection weight $\Omega$.  The time delays between vertices are a function of their positions in space.  The magnitude of the connection weights, $\Omega_{ij}$, are estimated based from vertices' known positions and the observed dynamics.}
\label{fig:framework}
\end{center}
\end{figure}

\begin{figure}[htbp]
\begin{tabular}{c|c}
\begin{tikzpicture}
	\node [draw,circle] (1) at (0,3) { $\yb_{1}(t)$};
	\node [draw,circle] (2) at (3,3) { $\yb_{2}(t)$};
	\node [draw,circle] (3) at (6,4) { $\yb_{3}(t)$};
	\node [draw,circle] (4) at (2,1) { $\yb_{4}(t)$};
	\node [draw,circle] (5) at (6,1) { $\yb_{5}(t)$};
	\path[->]  
		(1) edge node[above] {\small \textcolor{blue}{$\textbf{3.6}$}} node[below] {\small \textcolor{dgreen}{$-1.0$}} (2)
		(1) edge [bend left] node[sloped,above] {\small \textcolor{blue}{$\textbf{7.4}$}} node[sloped,below] {\small \textcolor{dgreen}{$1.0$}} (3)
		(3) edge node[sloped,above] {\small \textcolor{blue}{$\textbf{3.8}$}} node[sloped,below] {\small \textcolor{dgreen}{$1.0$}} (2)
		(2) edge node[sloped,above] {\small \textcolor{blue}{$\textbf{2.6}$}} node[sloped,below] {\small \textcolor{dgreen}{$0.7$}} (4)
		(4) edge node[sloped,above] {\small \textcolor{blue}{$\textbf{4.8}$}} node[sloped,below] {\small \textcolor{dgreen}{$-2.0$}} (5)
		(4) edge node[sloped,above] {\small \textcolor{blue}{$\textbf{3.4}$}} node[sloped,below] {\small \textcolor{dgreen}{$-2.0$}} (1)
		(3) edge node[sloped,above] {\small \textcolor{blue}{$\textbf{3.6}$}} node[sloped,below] {\small \textcolor{dgreen}{$1.0$}} (5)
		(5) edge node[sloped,above] {\small \textcolor{blue}{$\textbf{4.4}$}} node[sloped,below] {\small \textcolor{dgreen}{$-1.0$}} (2);
\end{tikzpicture} & \includegraphics[scale=.45]{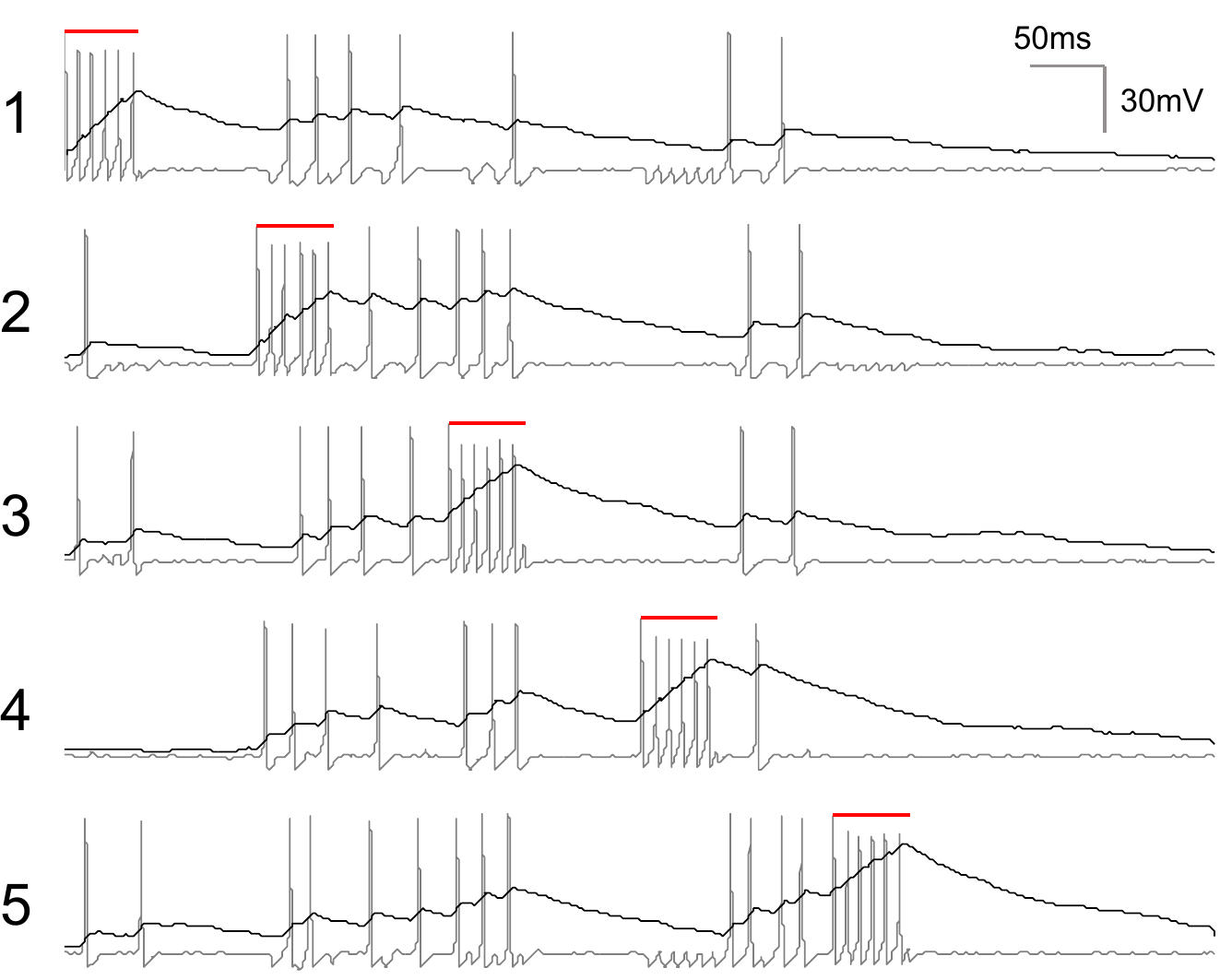}
\end{tabular}
\begin{center}
\caption{Signaling dynamics of the network from figure \ref{fig:framework}.  A Hodgkin-Huxley model is used for single-cell dynamics in a one second simulation.  The delays (written in \textcolor{blue}{blue}) are in milliseconds and are based on the cartesian distances between cells.  Functional weights, shown as relative strengths in \textcolor{dgreen}{green}, are chosen arbitrarily for the purposes of this example.  The right panel shows the network dynamics for each of the five cells following a sequential pulse stimulation (\textcolor{red}{red bar}) at each cell. Experimentally observable calcium traces are shown as solid black lines on an arbitrary vertical scale. Voltage, a hidden variable, is shown as a gray line and constitute the neurons' action potentials.}
\label{fig:signaling}
\end{center}
\end{figure}
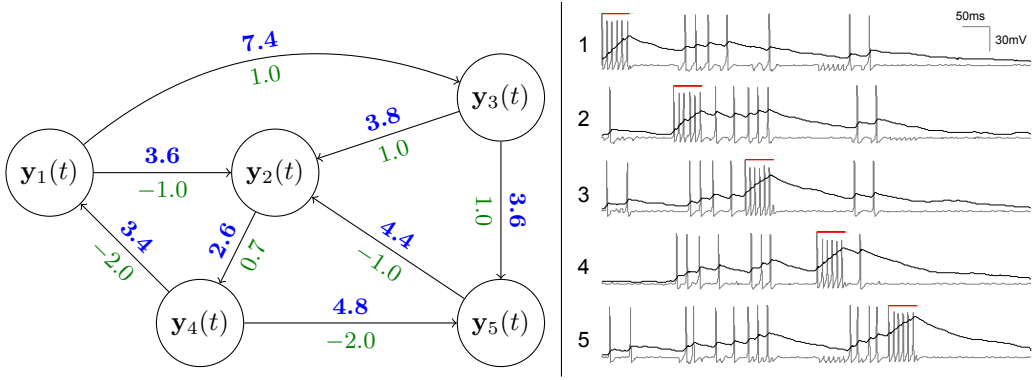

The framework can accommodate essentially all models of both neuronal and astrocytic dynamics. Independent of the specifics of any single cell model chosen, the framework provides a compact mathematical structure that quantitatively describes signaling and information propagation and flow in geometrically defined networks. The geometry and physical connectivity topology of the network can be simulated (e.g. random, scale free, or small world) or measured from experimental data such using methods such as optical imaging. Regardless of how one choses to set up the network, the framework provides a description of information flow through the network given knowledge of temporal signaling delays and chosen single cell models, or can be used to identify and map unknown functional connectivities and parameters in real neural circuits and networks. In all cases, the framework is able to provide an estimate of the complete description of the functional network and the interaction between all observable and hidden state variables and parameters. Figure \ref{fig:framework} illustrates a simple five vertex example that summarizes everything that is needed to describe the functional dynamics of information flow through the network. Figure \ref{fig:signaling} provides a specific example of the network from figure \ref{fig:framework} using a Hodgkin-Huxley model and simulating one second worth of data. Note how the framework provides experimentally measurable variables (calcium and membrane voltage) for every cell in the network in the temporal sequence dictated by the geometry and connectivity of the network. 

\subsection{Individual Cell Dynamics}\label{sec:indyn}
In this section we discuss how neuronal models of single cell dynamics, synaptic connections, plasticity, and observation fit within the framework. We begin by showing how to construct the state transition of an isolated (unconnected) vertex, and then build the full transition function by connecting multiple vertices into a network.  In an isolated vertex case with no incoming connections, the state transition reduces to
\begin{equation}\label{eq:basicstate}
\yb(t+\Delta t)=\Hb(\yb(t),\mathbf{u}(t),\Theta)
\end{equation}
This generic form encompasses neuronal models described in differential equation form, as well as those with a state reset based on some threshold value.  Most neuronal models are expressed in differential form as
\begin{equation}\label{eq:difform}
\frac{d\yb}{dt}=h(\yb(t),\mathbf{u}(t),\Theta)
\end{equation}
Converting \ref{eq:difform} into the state transition form given by \ref{eq:basicstate} is a matter of numerical integration with an integration method of choice.  Using Euler's method, for example, the state transition function $\Hb_{j}$ of the system in \ref{eq:difform} becomes
\begin{equation}
\Hb_{j}(\yb_{j}(t),\ub_{j}(t),\Theta_{j})=\yb_{j}(\Theta,t)+\Delta t\cdot h(\yb_{j}(t),\ub_{j}(t),\Theta_{j})
\end{equation}
Here we used the Euler method of integration for its simplicity and clarity, but other, but more complex integration methods like trapezoidal or Runge-Kutta can also be used to generate the next time step from the current step.

As an example, consider the simple Fitzhugh-Nagumo oscillator used to model neurons. In its differential form the model is given by the pair of equations
\begin{align}
\frac{dV(t)}{dt}&=aV(t)-bV(t)^3-cW(t)+S(t)+U(t)\\
\frac{dW(t)}{dt}&=e(V(t)+f-gW(t))\nonumber
\end{align}
The state set $\yb(t)$ is comprised of two state variables $\yb_{j}(t)=\{V_{j}(t),W_{j}(t)\}$. The experimental control set is composed of only one variable, affecting the $V(t)$ state variable, so $\ub_{j}(t)=\{U_{j}(t)\}$.  This system can be expressed in state transition form as $\Hb_{j}=\{H_{V,j},H_{W,j}\}$, with
\begin{align}\label{eq:fnstate}
H_{V,j}(t)&=V_{j}(t)+\Delta t \cdot \big( aV_{j}(t)-bV_{j}(t)^3-cW_{j}(t)+U_{j}(t) \big) \\ \nonumber
H_{W,j}(t)&=W_{j}(t)+\Delta t \cdot \big( e(V_{j}(t)+f-gW_{j}(t)) \big)
\end{align}
Note that we index the state transition functions based on the state variables they operate on; for example, $H_{V}(t)$ advances $V(t)$.  The parameter set for this system is composed of the parameters for each of the state transition equations in \ref{eq:fnstate}:
\begin{align}
\Theta_{V,j}&=\{a,b,c\} \nonumber \\ \nonumber
\Theta_{W,j}&=\{e,f,g\} \\ 
\Theta_{j}&=\Theta_{1,j}\cup\Theta_{2,j}
\end{align}
This system has two state variables and six parameters.  

Another class of neuronal models are those with a hard reset.  These models are also described in differential equation form, but contain a hard reset when a state variable reaches a certain value.  As an example, consider the Izhikevitch simple model, written in differential form as
\begin{align}
\frac{dV(t)}{dt}&=\frac{1}{C}\Big[k(V(t)-V_r)(V(t)-V_t)-W(t)+S(t)+U(t)\Big]\\ \nonumber
\frac{dW(t)}{dt}&=a\big(b(V(t)-V_r)-W(t)\big)\\ 
& \left. \begin{array}{rl}
V(t+)&=c\\ \nonumber
W(t+)&=W(t)+d 
\end{array} \right\} \text {  if }V(t)>V_{peak}\text{ (spike event)} 
\end{align}
This model consists of a voltage and amplifying currents ($V(t)$) and a resonant gating variable ($W(t)$).  The system has up to 9 parameters, and resets both state variables when a certain voltage threshold ($V_{peak}$ parameter) is surpassed. Mapped onto our framework, the model and its transition functions are
\begin{align}\label{eq:izsimple}
H_{V,j}&=\left\{
\begin{array}{rl}
	V_{j}(t)+\Delta t\cdot\frac{1}{C}\Big[k(V_{j}(t)-V_r)(V_{j}(t)-V_t)-W_{j}(t)+U_{j}(t)\Big] & \text{if }V_{j}(t)<V_{peak} \\
	V_{reset} & \text{otherwise} \end{array} \right .  \\
H_{W,j}&=\left\{
\begin{array}{rl}
	W_{j}(t)+\Delta t\cdot\big[r\big(b(V_{j}(t)-V_r)-W_{j}(t)\big)\big] & \text{if }V_{j}<V_{peak}\\
	W_{j}(t)+d & \text{otherwise} \end{array} \right . \nonumber
\end{align}
Here, the parameter set is $\Theta_{j}=\{C,k,V_{r},V_{t},V_{reset},V_{peak}\}\cup\{r,b,d,V_{r},V_{peak}\}$, and just like the Fitzhugh-Nagumo model, $\yb_{j}(t)=\{V_{j}(t),W_{j}(t)\}$,  $\ub_{j}(t)=\{U_{j}(t)\}$ and $\Hb_{j}=\{H_{V,j},H_{W,j}\}$. By changing the parameter values of the individual models, different classes of neurons can be simulated with the same transition function.  


Similarly, any model can be accommodated and fit into the framework, from the simplest to the most complex. Traditionally, all neuronal models have membrane voltage as a state variable and propagate a discrete signal in the form of an action potential when the membrane voltage rises past some threshold value at the axon hillock in response to depolarizing and hyperpolarizing currents in dendrites mediated by spatial and temporal summation of presynaptic currents.  The simplest neuronal model, the leaky integrate and fire (LIF) has voltage as a single state variable that decays to a target value and is perturbed by incoming currents.  If the voltage rises past a threshold value it is reset at the next time step to a reset value. One of the most complex and realistic single-cell models is the Hodgkin-Huxley (HH) model which relies on four state variables $\{v,m,n,h\}$ to describe the dynamics responsible for the generation of action potentials. The number of parameters increases with the number of state variables, from 4 in the LIF model to 22 for the HH model. Additionally, the required time step is shorter for HH models, being on the order of 0.03 milliseconds compared to roughly 5 milliseconds for the LIF model. The increased number of state variables and parameters along with shorter time steps puts a significant computational burden on any simulation or mapping algorithm.  The question of which model and how much complexity is required to best describe real-world data is not trivial and depends on the purpose and intent of the modeling.  

Although considerably more limited than the number of existing and studied neuronal dynamic models, there are a few astrocyte dynamic models that emphasize differing aspects and processes of astrocyte signaling \cite{Stamatakis:2006p1199, Nadkarni:2008p3684, DePitta:2009p1741, Bennett:2005p990, Lavrentovich:2008p1531, Macdonald:2008p1524}. Astrocyte models are expressed in differential forms similar to equation \ref{eq:difform}. Further research into astrocytic models is important though because astrocytes have been shown to play a direct role in the bidirectional communication between themselves and neurons via intracellular calcium transients and intercellular calcium waves under controlled experimental conditions  \cite{Fields:2009p556,Agulhon:2008p1532,Verkhratsky:2006p1527,Coco:2003p2473,Macdonald:2008p1524,Scemes:2006p1272} and more recently physiologic conditions in the neural retina \cite{KurthNelson:2009p4229} and cerebellum \cite{Hoogland:2009p4226}. Pathophysiologically intercellular calcium waves in astrocytes independent of neuronal hyperactivity have recently been shown to occur spontaneously \textit{in vivo} in the APP/PS1 transgenic mouse model of AlzheimerÕs disease \cite{Kuchibhotla:2009p1752};  and amyloid beta has been shown to be sufficient to trigger complex temporally delayed intercellular calcium waves in isolated astrocyte networks \cite{Chow:2009p6436}.

The state transition framework handles all single-cell dynamic models, as well as heterogenous systems of different cell types, either by different parameter sets or state transition function sets or both. Whatever the dynamics of individual neurons or astrocytes, all perform the same general task whereby processes and inputs generate outputs to other cells in a connected network.

\subsection{Cellular Network Signaling}\label{sec:netsig}
There are three components to cellular signaling: how long it takes for information from one cell to reach another, what are the effects of one cell on another and how do those effects change through time given the relative dynamics of the two connected cells. In this section we describe how signaling delays, functional connectivity and plasticity are accomodated by the framework.

\subsubsection{Signaling Delays}
Neurons and astrocytes form signaling networks that pass and process information between cells, and the state transition function must be extended to include signal propagation between functionally connected cells.  In biological cellular networks, signal propagation occurs at a finite and relatively slow speed (i.e. compared to electronic circuit networks).  Information in cell networks propagates on the order of microns per second for astrocytes to meters per second for myelinated axons in neurons.  Thus the influence of the dynamics of one cell is felt by another cell after some delay $\tau_{ij}$.  While the general form for the delay is given by \ref{eq:gendelay} above, the simplest form it can take is when signaling is a geodesic between vertices (or between the centers of cell bodies or centers of other cellular compartments as needed in morphologic models) and the transmission speed $s$ is constant: 
\begin{equation}\label{eq:delaydist}
\tau_{ij}=\frac{\| \xb_i - \xb_j\|}{s}
\end{equation} 

i.e. 
\begin{equation}
\tau_{ij}=D(\Gamma,\xb_i,\xb_j) \text{, } \Gamma_{ij} = \{s\} \text{ (\textit{c.f.}} \ref{eq:gendelay} \text{ above)} \nonumber
\end{equation}

Here the delay is simply the Euclidean distance between the cell centers divided by the transmission speed. For a diffusive network, as is the case with astrocytes, delays are proportional to the square of the distance between vertices. A more complex delay function may take into account knowledge about the particular physiology of the network, curved paths between cells, non-uniform speeds, etc.  The dependency of the framework on the delays is critical to its ability to describe how and when information within the network is processed, ultimately to a significant degree dictating the intercellular dynamics of the overall neural circuit or network. 

\begin{figure}[htbp]
\begin{center}
\includegraphics[scale=.7]{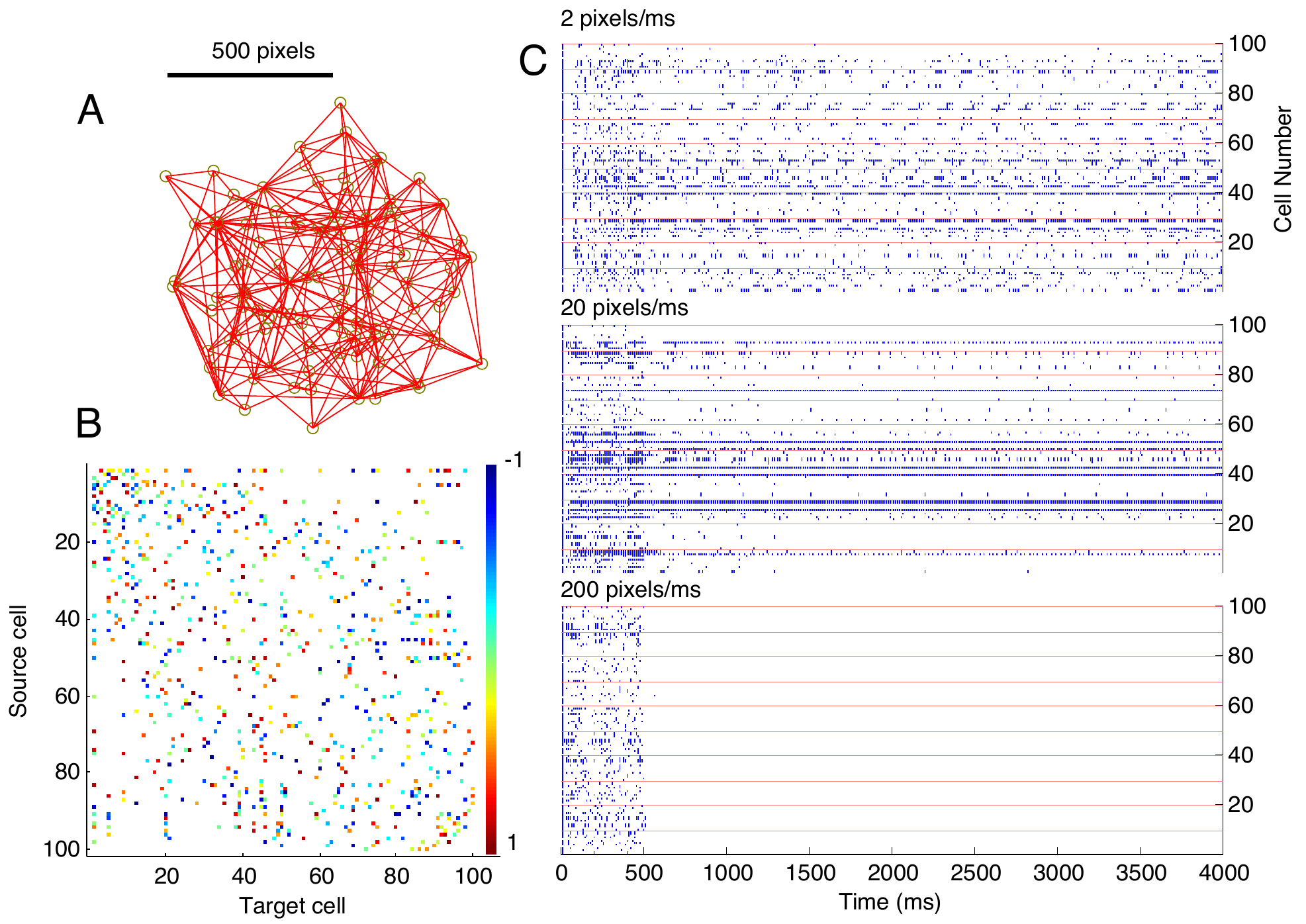}
\caption{{Effects of signaling speed on network dynamics.}The network with spatial locations and physical connections shown in panel \textbf{A}, is assigned random weights uniformly distributed between -1 and 1 on each physical edge, panel \textbf{B}. An Izhikevitch simple model of bursting neurons was used to model the individual vertex dynamics. \textbf{C.} By varying the speed of signal propagation, the delay distributions are scaled, having a substantial impact on the spike dynamics (see text).}\label{fig:delays} 
\end{center}
\end{figure}

Figure \ref{fig:delays} illustrates the dependency of network dynamics on signaling speed and delay times in a 100-vertex three dimensional network by varying the intercellular signaling speed, with everything else, including its geometry (i.e. its physical connectivity), the functional connectivity and input stimulus, remaining the same. We stimulated all cells with 500 ms of depolarizing current.  The delays are inversely proportional to the signal propagation speed.  We illustrate the effects of three signaling speeds, 2, 20, and 200 pixels/ms. At the lowest signaling speed, 2 pixels/ms, a low-frequency periodic activity was produced that qualitatively resembles a central pattern generator.  At a speed of 20 pixels/ms fewer cells exhibited low frequency oscillations, and signaling became more sporadic.  For both the 2 and 20 pixel/ms propagation speeds however, signaling continued past the period of stimulation. At a propagation speed of 200 pixels/ms however, there was no signaling past the stimulus period. In fully recurrent networks such as the one illustrated here, delays serve as a form of signal storage, essentially giving cells time to recover from a refractory period between activations, which in turn maintains recurrent signaling propagation well beyond an initial stimulus. For some appropriate range of signaling speeds, and therefore delay times, this recurrent signaling can settle into a repeatable pattern. If however, the signaling speeds are too fast, incoming signaling from upstream cells never have an opportunity to activate downstream cells because they are still refractory and do not respond. This leads to signaling in the network quickly dying away and not being sustained without it being driven by an external stimulus, as is the case with speeds of 200 pixels/ms in this example. A full discussion of the dependency of the network dynamics on the variables that govern it is very involved and beyond the scope of this paper. However, this example serves to illustrate that along with a network's connectivity topology and individual vertex dynamics, signaling delays play a crucial role in its overall response and dynamics and must be part of any network simulation or modeling framework that attempts to capture the inherent behaviors of neurobiological networks. 

\subsubsection{Functional Connectivity}
The other component of signaling is the functional connectivity of the network, or how the state of one vertex influences the state of another. In equation \ref{eq:Yjset} the set $\mathbf{Y}_{j}$ collects all the states of all the vertices in the network except $j$, delayed by a time value relative to $j$.  When $\mathbf{Y}_{j}$ is passed into the transition function $\Hb_{j}$, the information contained in every state of every other vertex is made available to all state variables in vertex $j$.  Within the framework, this is the broadest possible scope of connectivity, though in practice typically information from one variable affects one or more variables in another vertex.

As an example, consider a neuron and its pre synaptic (chemical) inputs that induce post synaptic currents.  The post synaptic current (PSC) is modeled with additional state variables in the neuron's state vector $\yb_j(t)$.  There are two PSC models in wide use which we have tested within the framework, although it is in no way limited to these two examples.  In general,  a pre synaptic neuron causes a PSC in a post-synaptic neuron that ultimately affects membrane voltage. Modeling the effect of an incoming signal on a target cell is key to establishing connectivity based on the observed cell dynamics.  
Post synaptic current is represented as a function of the form
\begin{align*}
s(t)=g_{max} \cdot r(t) \cdot (V(t)-E_{rev})
\end{align*}
where the resultant signals $s_j(t)$ are summed and passed to the voltage state variable in connected vertex $i$.  $g_{max}$ is the maximum conductance, and in the case of a specific synapse, can be expressed as a product of the maximum allowable conductance and the (instantaneous) functional connection weight $\omega_{ij}$. $r(t)$ describes the time course of the current, and is generally one of two forms, depending on the neuron type and neurotransmitter release \cite{DESTEXHE:1998p7063,DESTEXHE:1994p7058}:
\begin{align*}
r(t)&=e^{-at}&\text{ simple exponential}\\
r(t)&=at \cdot e^{-at}&\text{ $\alpha$-function}
\end{align*}
where $a$ is the time decay constant. In both cases, time starts at the moment of activation, in this case the time of the activation of the pre synaptic cell plus the delay to the post synaptic cell, $\tau_{ij}$.  For either case, the expressions for $r(t)$ can be written as linear differential systems, with the spike as the impulse.  For the simple exponential the differential equation is
\begin{align}\label{eq:simplepsc}
\frac{dr_j(t)}{dt}&=-ar_j(t)\nonumber \\
r_j(t+)&=r_j(t)+\omega_{ij}\quad \text{ upon arrival of spike event from vertex $i$}
\end{align}
For the $\alpha$-function, another state variable $p(t)$ is used:
\begin{align}\label{eq:alphapsc}
\frac{dr_j(t)}{dt}&=p_j(t) \nonumber \\
\frac{dp_j(t)}{dt}&=-a^2r_j(t)-2ap_j(t) \nonumber \\
p_j(t+)&=p_j(t)+\omega_{ij}\quad \text{ upon arrival of spike event from vertex $i$}
\end{align}
Shifting the arrival delays is simply a matter of shifting the spike detection function of the pre synaptic neuron,  so the arrival time from vertex $i$ to vertex $j$ is effectively the time shifted function of the voltage of $i:V_i(t-\tau_{ij})$. The synaptic current is a decreasing exponential with rate $a$, incremented by a weight value $\omega_{ij}$ upon arrival of a spike occuring $\tau_{ij}$ time units ago at another cell $i$.

For example, extending the state set for the Izhikevitch model given by \ref{eq:izsimple} for an arbitrary vertex to include an $\alpha$-function PSC model of post synaptic currents produces $\yb_{j}(t)=\{V_{j}(t),W_{j}(t),r_{j}(t),p_{j}(t)\}$ and $\Hb_{j}=\{H_{V,j},H_{W,j},H_{r,j},H_{p,j}\}$ where,
\begin{align}\label{eq:izpsc}
H_{V,j}&=\left\{
\begin{array}{rl}
	V_{j}(t)+\Delta t\cdot\frac{1}{C}\Big[k(V_{j}(t)-V_r)(V_{j}(t)-V_t)-W_{j}(t)+r(t)+U_{j}(t)\Big] & \text{if }V_{j}(t)<V_{peak} \\
	V_{reset} & \text{otherwise} \end{array} \right .  \\
H_{W,j}&=\left\{
\begin{array}{rl}
	W_{j}(t)+\Delta t\cdot\big[r\big(b(V_{j}(t)-V_r)-W_{j}(t)\big)\big] & \text{if }V_{j}<V_{peak}\\
	W_{j}(t)+d & \text{otherwise} \end{array} \right . \nonumber \\
H_{r,j}&=r_{j}(t)+\Delta t \cdot p_{j}(t)  \\
H_{p,j}&=p_{j}(t)+\Delta t \cdot \left(-a^2r_j(t)-2ap_j(t)+\sum_{k=1,k\ne j}^{J}\omega_{kj}\text{spd}(V_{k}(t-\tau_{kj}))\right)
\end{align}
The spike detection function is given by
\begin{equation}\label{eq:spd}
\text{spd}\left(V(t)\right)=\left\{ \begin{array}{rl} 1 & \text{if }V(t)\ge V_{peak} \\ 
0 & \text{otherwise} \end{array} \right . 
\end{equation}
Here, the spike inputs from other neurons are passed into the state variable $p(t)$ with transition function $H_{p,j}$, integrated by variable $r(t)$, and passed into the voltage variable $V(t)$.  The functional weights only operate on the voltage spikes from other neurons, so only information from the voltage state variable $V(t)$ is passed to other neurons.  A similar set can be constructed with the PSC models in \ref{eq:simplepsc} .

\subsubsection{Plasticity}\label{sec:plasticity}
Up to this point we have described models with fixed functional connectivity strengths.  This assumption is valid for networks observed over short periods where connection strengths can be assumed to be constant for the purposes of mapping or simulating since the plasticity mechanisms that modulate connection strengths operate on longer time scales.  If the weights change as a function of the activities of the cells it connects, the framework can be used to modulate connective strengths (equations \ref{eq:omegatrans} and \ref{eq:Gparams}).  Just as with single cell dynamics and network connectivity, there are many models of plasticity and we will not attempt to list or review them all here.  Rather, we describe how a simple spike-time dependent plasticity model reviewed by Bi and Poo in \cite{Bi:2001p6577} that incorporate long-term potentiation (LTP) and long-term depression (LTD) can be easily implemented within the proposed framework.

In equation \ref{eq:omegatrans} we described the functional strength or weight $\Omega_{ij}$ with transition function $\mathbf{G}_{ij}$ analagous to the state set for individual neurons.  When the weight is constant, the set contains only one variable so $\Omega_{ij}=\{\omega_{ij}\}$ and there is no temporal change in $\omega_{ij}$ and no transition function or parameters.  However, if the weight is modulated by the activitiy of cell $i$ on $j$ (the connection is directional, so $\omega_{ij}$ modulates information flowing from $i$ to $j$), then the states $\yb_{i}(t-\tau_{ij})$ and $\yb_{j}(t)$ affect how $\Omega_{ij}(t)$ changes in time.

The neuronal plasticity model in \cite{Bi:2001p6577} describes strengthening and weakening of synaptic conductance based on the timing of spikes in pre and post synaptic neuronal spiking.  If a post synaptic spike occurs immediately after a pre synaptic spike, the synaptic conductance is increased and the functional connection is effectively strengthened.  If the post synaptic neuron spikes before the pre synaptic neurons, the connection is weakened.  Other conditions like spike coincidence or long times between the spikes of pre and post synaptic neurons have no affect on the synaptic conductance.

To incorporate this model into our framework we first augment the set of cell states with another state variable $s_{j}(t)$, that stores the time from the last spike.  This state variable has transition function $H_{s,j}$:
\begin{equation}
H_{s,j}=(1-\text{spd}(V_{j}(t))\cdot(s_{j}(t)+\Delta t)
\end{equation}

The model in \cite{Bi:2001p6577} uses two exponential curves to describe changes in synaptic conductance based on the spike times.  By defining the function $q_{ij}(t)$ as
\[ q_{ij}(t)=(s_{i}(t-\tau_{ij})-s_{j}(t))\cdot\text{max}(\text{spd}(V_{i}(t-\tau_{ij}),\text{spd}(V_{j}(t)) \]
we can reconstruct the plasticity model within the framework as
\[ \frac{d\omega_{ij}(t)}{dt}=\alpha\cdot\text{sign}(q_{ij}(t))\cdot\text{exp}\left(-\beta|q_{ij}(t)|\right) \]
Thus, the transition function for dynamic synaptic weights is
\begin{equation}
G_{ij}=\omega_{ij}(t)+\Delta t\cdot\left(\alpha\cdot\text{sign}(q_{ij}(t))\cdot\text{exp}\left(-\beta|q_{ij}(t)|\right)\right)
\end{equation}
The parameter set for connection $ij$ is $\Lambda_{ij}=\{\alpha,\beta\}$.  This way, a static weight $\omega_{ij}$ can be converted into a dynamical one $\omega_{ij}(t)$, with behaviors governed by any arbitrary plasticity model and its associated parameter set $\Lambda_{ij}$.  It is important to note that while here we describe only one scalar weight between two vertices, the framework as defined can accomodate as many weights as there are state variables for a given vertex, thereby describing different classes of intercellular signaling between vertex pairs.  For example, in networks of neurons signaling may occur via gap junctional mediated electrical synapses in addition to chemical synapses, while in astrocyte networks intercellular signaling is typically mediated by diffusional processes (e.g. vesicularly released adenosine triphosphate, ATP).

\subsection{Experimental Observability Through Calcium Observation} \label{sec:observ}
Typically only one or a few of the state variables in the state vector are available for observation (i.e. are experimentally measurable). This is certainly the case with cellular networks, especially in neurons where voltage is measured as an indicator of signaling activity. But simultaneous voltage measurements are difficult for networks of many of neurons where the geometry of the network may be important to the analysis or interpretation of the data. While high density planar multi-electrode arrays can record from a few hundred cells at once, it is typically not possible to correlate recorded activity with the native geometry of the network. (There is one notable exception to this that we are aware of, which is the ganglion cell monolayer in the peripheral retina. Chichilnisky and colleagues are able to computationally infer the geometry and functional activity of these retinal ganglion cells due to their unique planar arrangement- see \cite{Pillow:2005p6755,Shlens:2008p6743,Chichilnisky:1999p6763}. However, in the brain and even in the retina where mutlilayered ganglion cells recieve incoming macular inputs it is not possible to do with electrode arrays.) Given the challenges associated with direct electrophysiological measurements of large neuronal ensembles, calcium fluorescence imaging has been used as an indirect measure of the dynamics of large neuronal netwokrs. The dynamics of calcium, often influenced by voltage spiking, are modeled as state variables and associated transition functions added to the state set. There are a few models describing the time course of calcium as it is driven by changes in membrane voltage, and attempts to develop more refined and robust calcium dynamics models specifically for the study neural microcircuits and networks is a very active area of research. Our intent here is not to describe on-going efforts or the state of the art but simply to illustrate the integration of one such model within our framework.

The simplest model of calcium dynamics can be expressed as a linear system, with a spike input
\begin{equation}\label{eq:casimple}
\tau_c\frac{dc(t)}{dt}=-\frac{1}{\tau_{c}}c(t)+\text{spd}(V(t))
\end{equation}
This model is integrated into the framework by the addition of another state variable $c(t)$ describing the cytosolic calcium concentration to the state set, with transition function:
\[ H_{c,j}=c_{j}(t)+\Delta t\left(-\frac{1}{\tau_{c}}c_{j}(t)+\text{spd}(V_{j}(t))\right) \]
This model has one parameter $\tau_c$ which describes the removal rate of calcium, after an input caused by a spike.  This is the simplest model used for calcium dynamics based on neuronal spiking and is often used to extract spikes from calcium \cite{Vogelstein:2009p4525}.  More complex and non-linear models of calcium dynamics have been developed and these can also readily be integrated within the framework .

Finally, there is the issue of observable variables.  In equation \ref{eq:obsfun} we defined an observable set $\mathbf{z}_j(t)$ as some function of the current state set $\yb_j(t)$.  When using fluorescent calcium indicators, the sole observation variable is the recorded intensity $I_j(t)$ for cell $j$ at a particular pixel reflecting some linear multiplier of the cytosolic calcium concentration at that point in the visual field, based on the dye loading in a cell, as well as the camera, microscope, and illumination setup.  Defining $\mathbf{z}_j(t)=\{I_{j}(t)\}$ with observation function $\mathbf{F}_{j}$,

\[ I_{j}(t)=\mathbf{F}_{j}(c_{j}(t),\mathbf{\Phi}_{j})=nc_{j}(t)+b+\mathcal{N}\left(0,\gamma^{2}\right) \]
The parameter set $\mathbf{\Phi}_{j}=\{n,b,\gamma\}$ represents the scaling, offset, and noise standard deviation of the observation function.  The function $\mathcal{N}(0,\gamma^{2})$ generates a normally distributed, random value with zero mean and $\gamma$ standard deviation.  The noise term models the type of frame-to-frame variation typically seen in the amplification of the CCD signal prior to digitization.  The size of $\gamma$ is proportional to the magnitude of the noise, itself affected by camera type and gain settings.

Since the framework is defined at cellular resolution, we are making the simplifying assumption that the recorded intensity represents the average intensity for the region of interest demarcating a specific cell $j$ within a larger field.  Additionally, if the camera records at a slower frequency than the transition dynamics $f_{camera}>1/\Delta t$, where $f_{camera}$ is the camera recording frequency and $\Delta t$ is the time incriment (\textit{c.f.} equation \ref{eq:Htrans}), one is averaging intensity values for the duration that the camera shutter stays open.  If this is the case then multiple sequential calcium concentration values would be averaged to produce a single intensity reading.

\subsection{GPU Implementation and Benchmarks}\label{sec:impl}

The practical application of the theoretical framework both for simulation and, as will be described in a subsequent paper, for mapping the unknown functional connectivity of experimentally observed cellular networks  necessitates its implementation in an appropriate computing environment. We have taken advantage of emerging high performance general purpose-graphics processing unit (GPU) parallel computing, although it can run as serial code on a normal central processing unit (CPU), which we have also tested.  Within the GPU environment, the code has been designed to run on nVIDIA graphics cards equipped with the CUDA interface (see http://www.nvidia.com/cuda).  In this way, we can parallelize vertex dynamics, signaling dynaimcs, and observation integrations over many processor cores, achieving significant speedup over CPU or cluster computations. The framework and associated single cell dynamic and network connectivity models have been coded as compact MATLAB-callable libraries.  All graphics user interface (GUI) and input/output (I/O) operations are handled through MATLAB and the code has been written in both MATLAB and plain C libraries that communicate through MATLAB.  The libraries offer direct control over all parameters for all vertices.  Using plain C language, any model that can be analytically described within the framework can be easily coded into a simulation library. From a practical experimental perspective GPU computation offers unprecedented scalability to larger systems with full access to all state variables and parameters, enabling rapid parallel simulation when the framework is run in the forward direction, and real-time dynamic mapping when the framework is applied to the inverse problem of mapping unknown functional connectivities of cellular neural networks. Speed and parallelization are critical for statistical simulation based identification methods such as particle filtering to operate in real time or near real time.

\begin{figure}[htbp]
\begin{center}
\includegraphics[scale=.5]{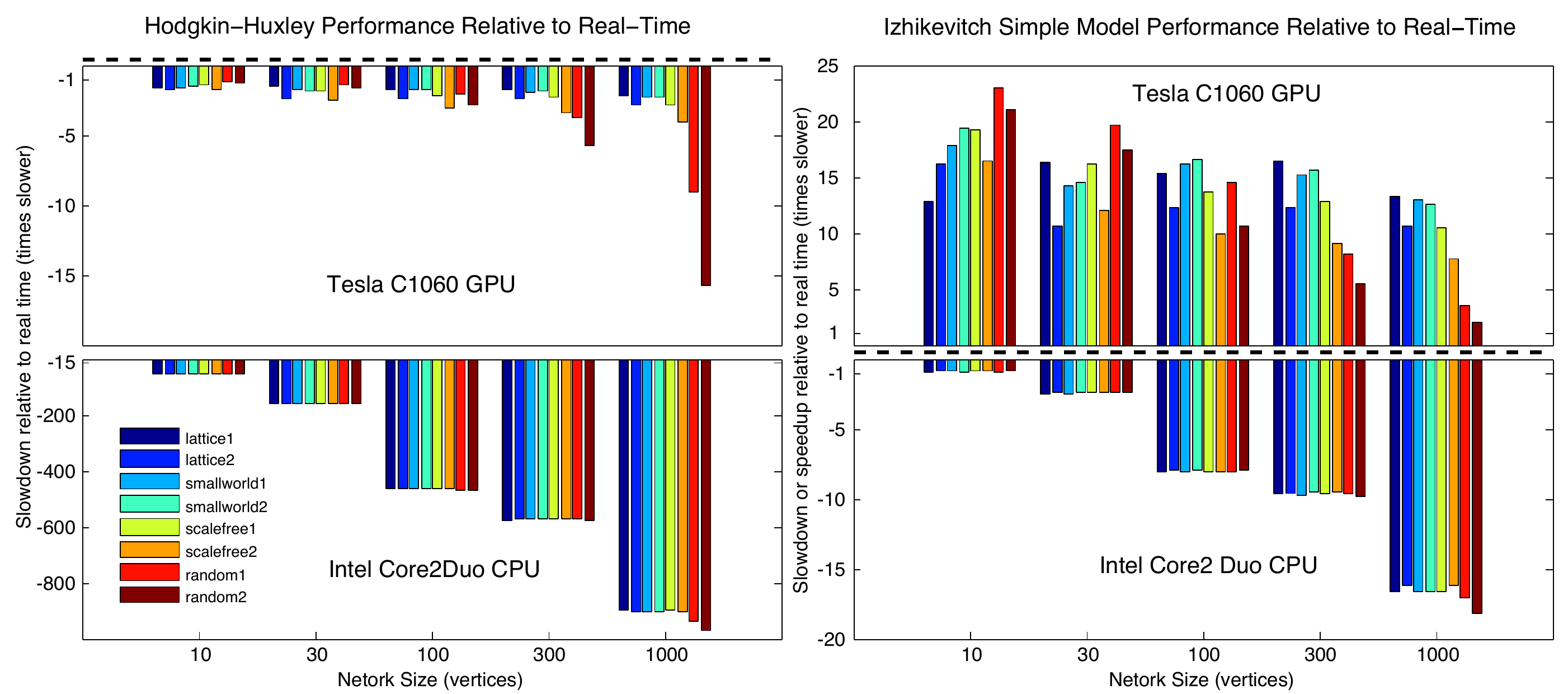}
\caption{CPU and GPU benchmark results for framework simulations using Hodgkin-Huxley and Izhikevitch models of single cell dynamics.  Two different network topologies for four different classes of networks were simulated (lattice, small world, scale free, and random), with network sizes (i.e. number of vertices) of 10, 30, 100, 300, and 1000 vertices simulated for each of the eight different networks. The top graphs show performance on a single nVIDIA Tesla C1060 GPU, expressed relative to the real time simulation period for each single cell dynamic model (see text), while the data for the bottom two graphs show the performance on a single core of the an Intel Core2Duo 2.5GHz processor. Negative values represent a slowdown relative to real time, while positive values represent a speedup. All forty networks were tested on both. See section \ref{sec:tests} below for details regarding the test networks and network classes. }
\label{fig:bench}
\end{center}
\end{figure}

Benchmarks in figure \ref{fig:bench} show the relative speeds of the CPU and GPU implementations of two different dynamic cell models within the framework, an Izhikevitch model and a Hodgkin-Huxley model, simulated in 40 test networks (see section \ref{sec:tests} below regarding the test networks).  The parallel GPU implementation performed anywhere from 8 to 200 times faster than the same code executed on a single CPU thread.  Performance was measured as a slowdown or speedup factor relative to real time. The Izhikevitch model had six state variables and a timestep of $\Delta t$ 1 millisecond. We simulated 10 seconds worth of data equivalent to 10,000 time steps. The more computationally intensive Hodgkin-Huxley model had eight state variables and a time step of 0.03 milliseconds.  One second worth of data was simulated for the HH model, corresponding to about 33,333 time steps. Benchmarks were calculated as the dimensionless ratio of the actual period of time simulated (i.e. real time) to the amount of physical computational time it took the GPU or CPU to carry out the simulations. Because of the parallel implementation on the GPU, the performance falloff was much slower than on the CPU with increasing network size.  In both CPU and GPU cases, performance decreased with increasing network density (total number of edges/vertices square), since in more dense networks more information is transferred between edges.  This is more apparent in the random networks that have higher density than other network classes, producing relatively slower processing speeds. In general, the constant time step in the framework makes parallelization easy on GPU architectures, delivering network simulation performance near or even faster than real-time. For the HH model, the GPU was able to carry out the computations in essentially real time for all of the networks tested except for the largest random networks, which were about 10 to15 times slower then real time for random networks with over a thousand vertices (upper left panel in figure \ref{fig:bench}). For computationally simpler models such as the Izhikevitch model the GPU computations were actually faster then the period being simulated (i.e. faster than real time), ranging from roughly 10 to 20 times faster for most networks and about real time for the largest random networks (upper right panel in figure \ref{fig:bench}). In contrast, CPU computations were always slower than real time, from 15 to 800 times slower for the HH model depending on the network class and size (lower left panel in figure \ref{fig:bench}) and from just under near real time for small 10 vertex networks to about 15 times slower for the largest random networks for the Izhikevitch model (lower right panel figure \ref{fig:bench}).  Additional GPU cards can further improve performance by splitting the task of advancing temporal cell dynamics;  however transfer of information between cells is still limited by memory and bus speeds, so dense networks will run slower than sparse networks. It is important to appreciate that the ability to carry out such forward simulations or to solve the inverse problem of mapping unknown functional connectivity topologies of networks in near real time or faster than real time using GPU computing is due to the mathematical construction of the framework and its efficient algorithmic implementation. It is impossible to do such network simulations or mappings in real time using biophysical compartmental simulation environments. The MATLAB and CUDA code for the framework are available for download from the authors' website (http://www.silva.ucsd.edu/Silva\_Lab/Links.html).

\subsection{Standardized Tests for Connectivity Estimation}\label{sec:tests}
Lastly, we propose a standardized basis test set to evaluate the effectiveness of mapping algorithms. A standard test set is well accepted in the field of non-linear optimization, providing a standard measure of different algorithms \cite{Hock:1980p4454}.  There are two benefits to having a standard set of networks to use for mapping.  First, multiple algorithms can be evaluated against the same network and model, providing relative performance benchmarks.  Second, data generated for a network using one dynamic model can be mapped using another dynamic model and comparisons can be made between the original topology and the mapped topology. This latter approach helps answer questions about model fitness, which are especially useful when trying to map data with multiple models or uncertainty in models. We emphasize that the test set we propose here should in no way be interpreted as implying that the full complexity and variability of real neural circuits and networks is captured or even described by the set. But we argue that any mathematical method or algorithm that claims to be able to deal with real cellular neural networks of any meaningful size (e.g. on the order of tens to hundreds of cells or larger) must at the very least be able to effectively and efficiently map functional networks derived by this test set, which offers a first order approximation to the dynamics and complexity displayed by such networks.

Here we offer the foundation for such a test set: The location of vertices in physical space and their physical connections based on different connective classes. The choice of dynamic model, parameters, and functional weights is left open and up to the discretion of the individual investigator, since they are specific to the network being studied and the mapping algorithms being designed, but can be directly implemented within the framework we have developed. Test networks vary in size from 10 to 1000 vertices, covering the range of cells that can be imaged simultaneously with fluorescence microscopy.  At the small end of the scale, networks on the order of 10 vertices is about the limit of existing connectivity estimation methods \cite{Makarov:2005p1762, Eldawlatly:2010p6733}.  The upper end of the scale at a 1000 vertices was chosen largely due to limits of computational power available at present.

Each graph is composed of $N$ interconnected vertices located in two or three dimensional physical space, with minimum distance constraints and other dynamic parameters as described below for each network class. The physical connectivity between vertices follow one of four different graph theoretical classifications, since there is no measurable and ``real" network spatial geometry defined in this case; this is also discussed below. Vertices were positioned in geometric space randomly, but with a prescribed minimum distance between neighbors. We developed a simple algorithm to populate a physical space with $N$ cells or nodes:
\begin{verbatim}
pick a random position for first cell X1
set i=2
while i<=N
     pick a random position Xi in space range
     set found=1
     for j=1 to i-1
          if cell Xi is less than d units from cell Xj
	     found=0
	     break for loop
	end if
     end for
     if found==1
          accept Xi as next cell
          increment i
     end if	
end while
\end{verbatim}

\begin{figure}[htbp]
\begin{center}
\includegraphics[scale=1.05]{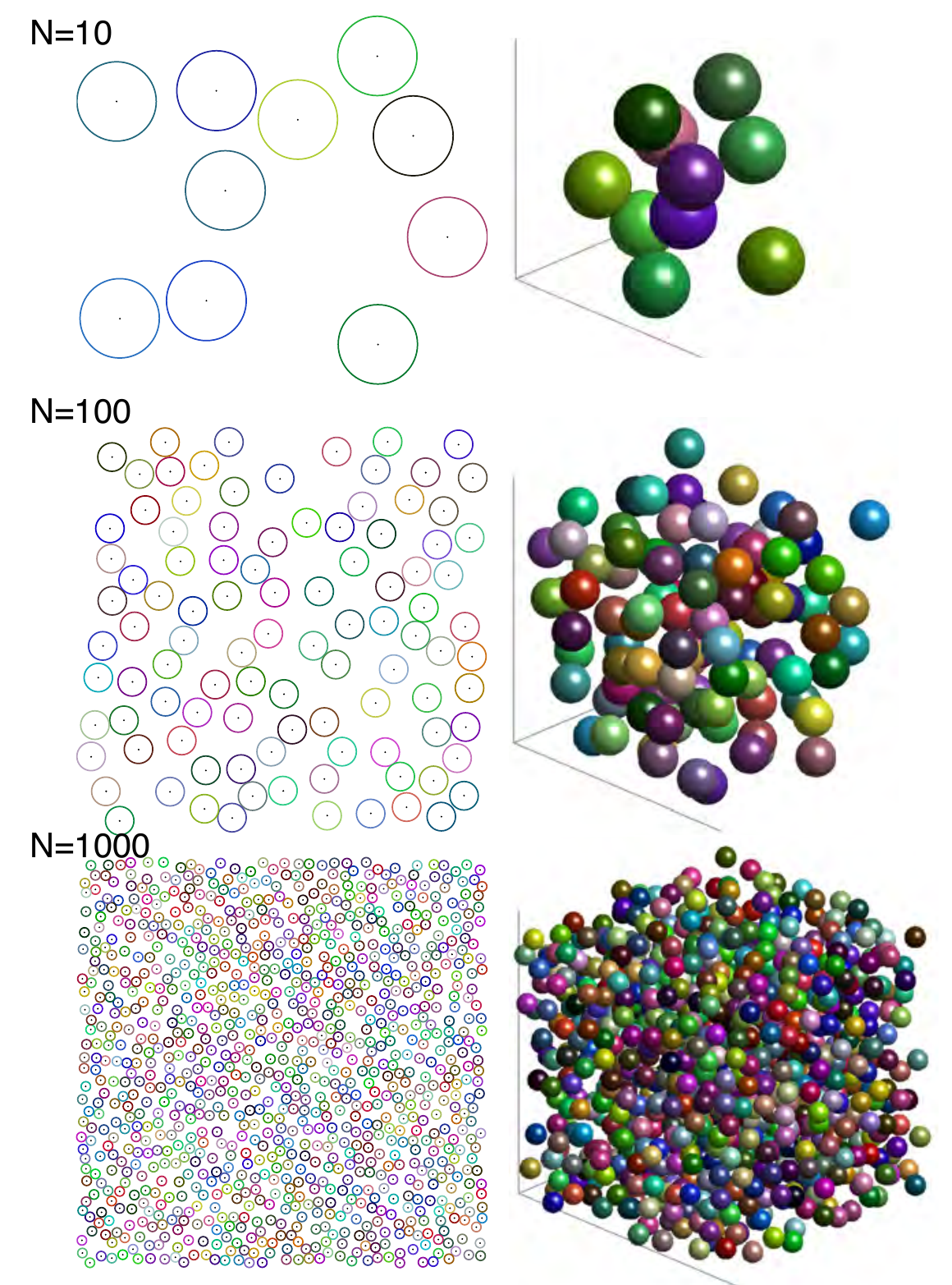}
\caption{Random placement of vertices with a minimum distance constraint.  Examples of groups of 10, 100, and 1000 vertices placed in two (left column) and three (right column) dimensions are shown here. Vertices were colored randomly for clarity. The minimum distance is a function of both the number of vertices and the spatial dimension.  For the same number of vertices, a two-dimensional network will have a smaller minimum distance and thus be packed tighter than a three-dimensional network.  See text for details.}
\label{fig:placement}
\end{center}
\end{figure}

The space range describes the dimensions of the space occupied by vertices in the network.  For a two dimensional network this could be a square area of 500 x 500 distance units.  Distance and position units are non-dimensional and can be scaled as needed.  The minimum distance between vertices is a parameter, but can also be expressed as a function of the number of vertices $N$ and the dimension of the space the graph occupies:
\[ d_{min}\propto LN^{-1/D} \]
Here the minimum distance $d_{min}$ is proportional to the length of the space $L$ times the number of vertices $N$ raised to the negative inverse of the dimension $D$.  For example, a three dimensional space of length $L=100$ could fit $N=1000$ vertices with minimum distance $d_{min}=100\times1000^{-1/3}$.  This implies a minimum distance of 10.  While this is the absolute minimum distance with cubic packing, when placing vertices at random this minimum distance is reduced to allow for some variability in placement.  Figure \ref{fig:placement} shows examples of vertex placement in two and three dimensions for three different size networks.
It is important to note that with this formula, the minimum distance can be prescribed for fractional dimensions, as may be the case for some neural tissues where cell arrangement is neither flat nor fully three dimensionally filling. Vertices are numbered from the center outwards, so mapping can be performed on a subset of vertices that interact with the complete network.  This is a more difficult case but a more realistic scenario, as vertices would be receiving inputs from unobserved vertices in real cellular networks due to experimentally limited windows of observability.

Once the vertices are placed in physical space, connections are made using established graph theoretic classes. We included lattice, small world, scale free, and random classes in the test set. For each connectivity class, we generated two networks of different edge densities, one with fewer and one with more edges. We chose these four classes because they represent the major graph theoretic topologies, but of course any other algorithmically defined class can be used. The physical connectivity of a graph intuitively represents a constrained phase space on which dynamic signals propagate in both space and time (i.e. the functional connectivity topology) as determined by the network signaling framework and chosen model of single cell dynamics. Specifically, we considered the following classes and specific parameters for each:\\

\noindent \emph{Lattice networks.}  This class of networks has only nearest neighbor connections with no long distance connections.  The number of nearest local connections or total number of edges can be specified before construction. In our case we limited connections for each vertex to its closest 3 and 8 neighbors.\\

\noindent \emph{Small world networks.}  This is a modification of a lattice network which includes a specified probability of long-range connections \cite{Barabasi:1999p4089}.  The probability of long range connections ranges between 0 (lattice network) and 1 (random network), but typical values are around 10 percent, meaning that 10 percent of all edges are randomly chosen.  In our case we built networks of 5 and 15 percent probability of random re-wiring.  \\

\noindent \emph{Scale free networks.} These networks follow a power law connection (edge) degree distribution, with many cells having few connections and few cells having many connections.  When positional aspects are taken into account, scale free networks take on some small world properties and are essentially scale free geometric graphs called apollonian networks  \cite{Andrade:2005p3037}.  \\

\noindent \emph{Random networks.}  The study of random graphs extends all the way back to the original work of Erd\"{o}s and R\'{e}nyi. In a random graph, a specified number of edges are placed between randomly chosen vertices, without regards for vertex position, which has no meaning. We built random networks of 10 and 20 percent densities, meaning about $(N^2*10\%)$ or $(N^2*20\%)$ number of edges where $N$ is the number of vertices in the network. A 100\% dense network connects every vertex to every other vertex.\\

\begin{figure}[htbp]
\begin{center}
\includegraphics[scale=.6]{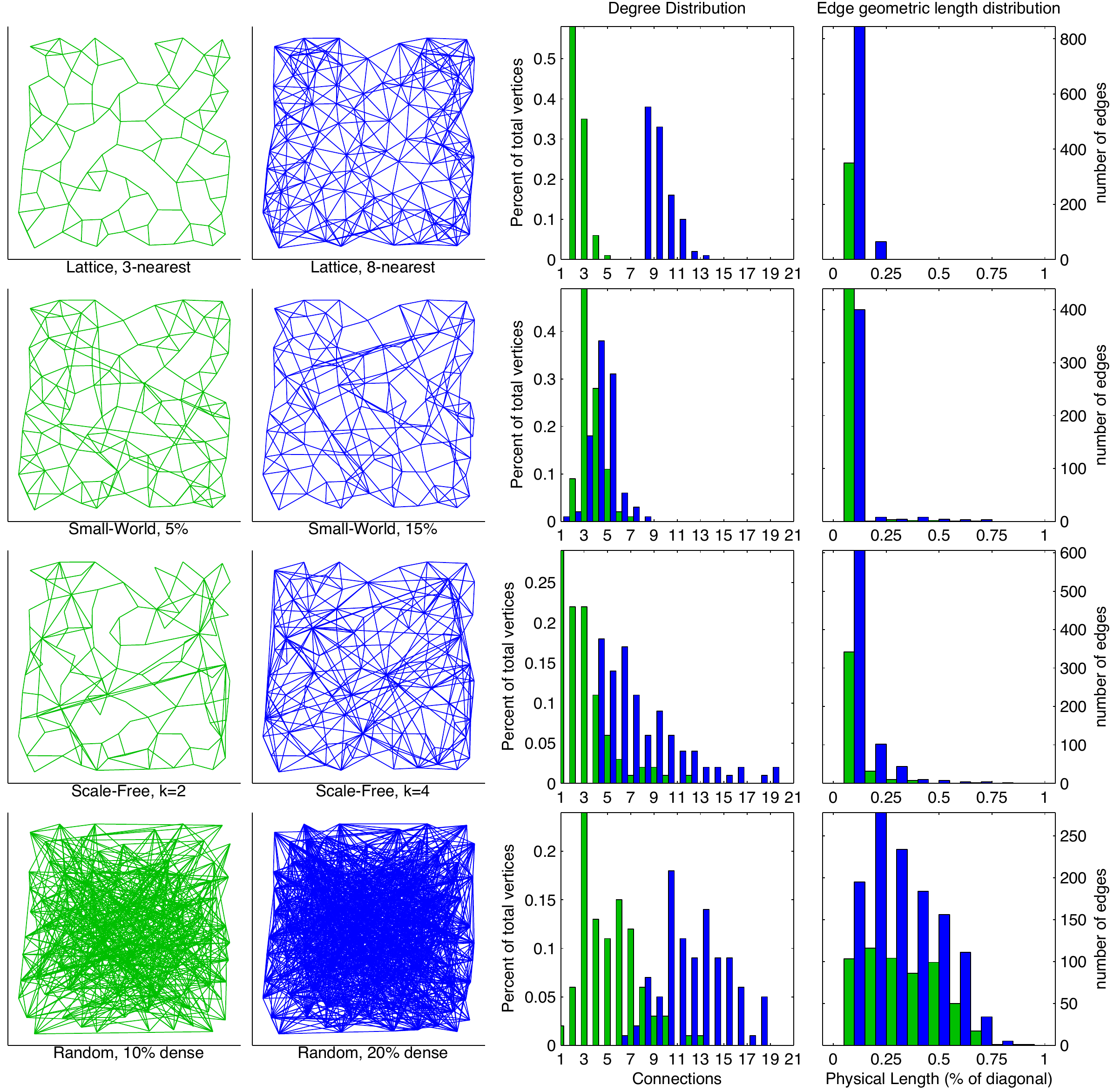}
\caption{Test network topologies for a 100-vertex two dimensional network. For each of the four topologies included in the test set two edge densities were considered, one with fewer edges or randomness (\textcolor{dgreen}{green}), and a more complex one with more edges or randomness (\textcolor{blue}{blue}; left two columns).  Parameter values for each network topology are shown below their respective graphs.  The two right-most columns plot the degree distribution (third column from the left) and the geometric length distribution (fourth column from the left) for each of the edge densities for each class.  The total number of edges for each network is the area under the edge distribution plots, while the total physical wiring length of the network is the area under the geometric length distribution plots.  The higher edge density networks (colored in blue) have both higher total number of edges and total wiring length than the lower edge density networks (colored in green).  For a constant signal propagation speed the wiring length distribution represents the delay time distribution between vertices.}
\label{fig:connectivity}
\end{center}
\end{figure}

\begin{figure}[htbp]
\begin{center}
\includegraphics[scale=.8]{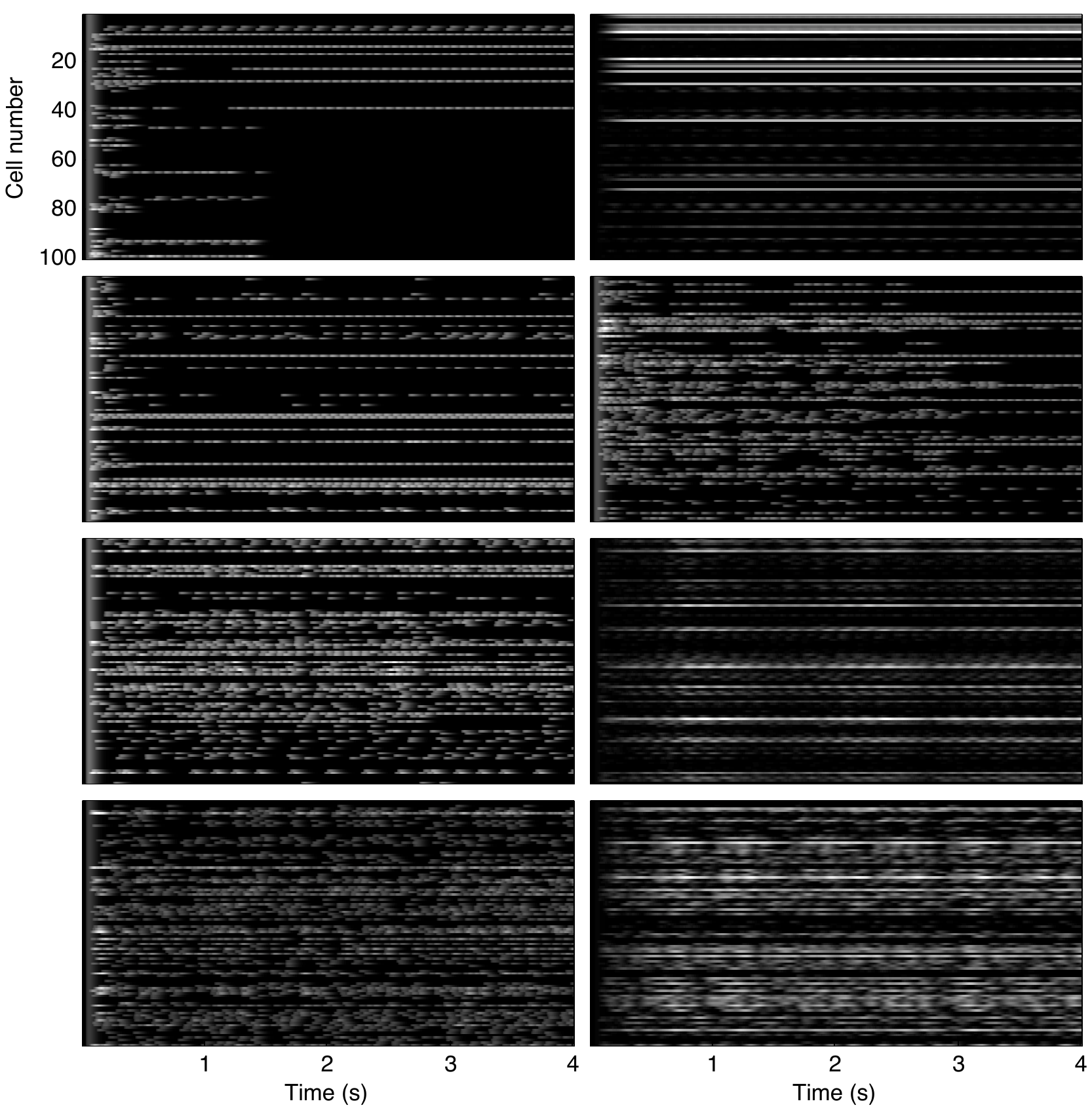}
\caption{Computed (i.e. simulated) dynamics of experimentally observable calcium signaling for each of the eight networks shown in figure \ref{fig:connectivity}, arranged in the same order.  Each network was simulated for four seconds using the Izhikevitch simple model, with a constant current input to generate activity.  The gray scale represents calcium activity with lighter shades representing the approximate instantaneous spike rate of individual neurons in the network.}
\label{fig:netsdyn}
\end{center}
\end{figure}

The different classes and densities are shown in figure \ref{fig:connectivity}, along with graph-theoretic statistics on connectivity and wiring lengths. The connections establish the physical connection between vertices or the edges along which functional connections are possible. The magnitude of the functional weights should be chosen based on the dynamic model and the specifics of the system studied.  A mapping algorithm must identify the functional connections without knowledge of the physical connectivity class. Delays are defined according to the cartesian distances between connected vertices.  Three-dimensional spaces generally have narrower and smaller distance distributions than two-dimensional packings. The formula presented in equation \ref{eq:delaydist} is used to establish delays, with the speed parameter chosen arbitrarily based on the system being studied. The complete test set is 80 networks, combinations of two dimensions, five sizes, and four connectivity classes, as outlined in table \ref{tab:modelspace}. Figure \ref{fig:netsdyn} shows an example of simulated calcium response raster plots for the networks shown in figure \ref{fig:connectivity}.  A mapping algorithm should be able to identify the dynamics parameters and the functional connectivity of each test network, given a chosen single cell dynamic model. Ultimately, the only kind of measured experimental data available to any such algorithm would be imaged calcium responses such as those simulated in figure  \ref{fig:netsdyn} or some equivalent data for another marker of functional cellular activity. These are the practical experimental constrains that any theoretical methods aimed at mapping functional activity in cellular neural circuits and networks with single cell resolution must face. The test networks are also available for download from the authors' website (http://www.silva.ucsd.edu/Silva\_Lab/Links.html). 

\begin{table}[htdp]
\begin{center}
\begin{tabular}{r|p{8cm}}
\textbf{Parameter} & \textbf{Range} \\
\hline \hline
Network Size $N$ & 10, 30, 100, 300, 1000 \\
\hline
Geometric Dimension & 2D, 3D (may even be fractal, i.e. 2.5D) \\
\hline
Connectivity Type  & lattice, small world, scale-free, random\\
\hline 
Edge Density & low, high \\
\hline \hline   
\end{tabular}
\end{center}
\caption{Range of parameters specified by the test set.  The current set is composed of eighty test networks span spanning ranges in network size, geometric dimension, connectivity type and edge density.  The choice of dynamical cell model, parameters, functional weights, observation variable, user inputs and noise levels are left to the individual user.}

\label{tab:modelspace}
\end{table}

\section{Discussion}
Within the study of networks, there are two opposing yet deeply interrelated processes: Simulation and estimation.  Simulation of networks deals with the forward problem of making predictions using an established model and measured parameters and connectivities.  The reverse problem, estimation or mapping, uses the result of actual collected data to infer, estimate, or map the parameters of a model or, for networks, functional connectivity.  The framework we introduce here can be used both for simulating signal propagation in physically realistic networks, and for the reverse process of estimating or mapping unknown functional connectivities of networks. The framework is bounded by a set of rules and constraints imposed by the experimental reality of cellular neurobiological methods: Complex non-linear dynamics, limited observability, noise and uncertainty, and experimental control.  It is designed around current observation and experimental capabilities, which are shifting from single neuron multiple trials, to multiple neuron single trial experiments \cite{QuianQuiroga:2009p3741}.

Within the framework, the mathematical construction for the dynamic model describes the time course of each vertex.  A general state transition representation encompasses different model types, from ordinary differential equations to state machines and Markov models, to simplified neuronal models like LIF and Izhikevitch. The choice of model will certainly affect the estimation of the dynamic parameters from the collected data;  how much the single-cell dynamic model affects the estimation of functional weights is is still an open question.  The proposed test set should help address this by simulating artificial data with one model and estimating with another.

Mapping a complete functional topology is ultimately a reverse process, and will involve some combination of estimation, filtering, and optimization.  While some approaches exist for estimating parameters and dynamics of single neurons \cite{Creveling:2008p3081} or small groups of neurons \cite{Makarov:2005p1762, Eichler:2003p3094, Eldawlatly:2010p6733}, mapping large networks within biologically realistic constraints remains a challenge, and we are still a long way from establishing a complete functional connectivity map of even simple processes and tasks. Indeed, from a neurophysiological perspective, it is not even entirely clear 'what' we should be mapping or how to properly interpret such data from the perspective of deciphering the neural code.  The proposed framework attempts to unify both theoretical and practical considerations as an ``open standard'' for the development of large scale functional topology reconstruction algorithms.

Since the goal of mapping is to identify both the dynamic parameters of individual vertices as well as the connectivity between vertices, a well-designed input control should be used to make the observation as informative as possible, provided it does not alter the parameters and connectivities of interest. Experimentally, input control can take on many forms. The dynamics of individual cells can be perturbed using methods such as optogenetics, pharmacologically using appropriate agonists and antagonists, and electrophysiology. For single cells there are different input functions that can be used and there are a few approaches describing input function design to extract the most amount of information \cite{Lewi:2009p7172, Benda:2007p1828}. At the network level, the set of input functions for each cell must be designed in parallel and coordinated with observed activity in order to provide the most amount of information to the mapping algorithm.

\subsection*{Acknowledgments}
This work was supported by grant RO1 NS054736 from the National Institute for Neurological Disorders and Stroke (NINDS) at the National Institutes of Health (NIH).\\

\bibliographystyle{abbrv}
\providecommand{\natexlab}[1]{#1}
\expandafter\ifx\csname urlstyle\endcsname\relax
  \providecommand{\doi}[1]{doi:\discretionary{}{}{}#1}\else
 \providecommand{\doi}{doi:\discretionary{}{}{}\begingroup
  \urlstyle{rm}\Url}\fi
 \bibliography{./mbcl}

\end{document}